# Quantitative analysis of atomic disorders in full-Heusler $Co_2FeSi$ alloy thin films using x-ray diffraction with Co-*Kα* and Cu-*Kα* sources

Yota Takamura[1,2,a)], Ryosho Nakane[3], and Satoshi Sugahara[1,2,4,b)]

[1] *Department of Electronics and Applied Physics, Tokyo Institute of Technology, Yokohama 226-8503, Japan*

[2] *Imaging Science and Engineering Laboratory, Tokyo Institute of Technology, Yokohama 226-8503, Japan*

[3] *Department of Electrical Engineering and Information Systems, The University of Tokyo, Tokyo 113-8656, Japan*

[4] *CREST, Japan Science and Technology Agency, Kawaguchi 332-0012, Japan*

________________________________________________________________

a) Electronic mail: yota@isl.titech.ac.jp

b) Electronic mail: sugahara@isl.titech.ac.jp

ABSTRACT

The authors developed a new analysis technique for atomic disorder structures in full-Heusler alloys using x-ray diffraction (XRD) with Co-$K\alpha$ and Cu-$K\alpha$ sources. The developed technique can quantitatively evaluate all the atomic disorders for the exchanges between $X$, $Y$, and $Z$ atoms in full-Heusler $X_2YZ$ alloys. In particular, the technique can treat the $DO3$ disorder that cannot be analyzed by ordinary Cu-$K\alpha$ XRD. By applying this technique to full-Heusler $Co_2FeSi$ alloy thin films formed by rapid thermal annealing (RTA), RTA-temperature ($T_A$) dependence of the atomic disorders was revealed. The site occupancies of Co, Fe, and Si atoms on their original sites were 98 %, 90 %, and 93 %, respectively, for the film formed at $T_A$ = 800 °C, indicating that the RTA-formed $Co_2FeSi$ film had the $L2_1$ structure with the extremely high degree of ordering.

BODY

Half-metallic ferromagnets (HMFs)[1,2] are very attractive for spintronic devices, such as magnetic tunnel junctions[3-5] and spin transistors[6,7], owing to their complete high spin polarization of 100 % at the Fermi energy. In particular, Co-based full-Heusler alloys, e.g., $Co_2FeSi$ (CFS)[8,9], $Co_2MnSi$ (CMS)[3], and $Co_2CrAl$ (CCA)[10], are promising as a HMF material, since they show the half-metallicity even at higher temperatures than room temperature. The half-metallicity of full-Heusler alloys is very sensitive to atomic disorders in their crystal structure (that is the $L2_1$ structure shown in Fig. 1(a))[11-13]. X-ray diffraction (XRD) technique has been commonly used to evaluate disordered structures[14-20]. However, XRD analysis with an ordinary Cu-$K\alpha$ source cannot distinguish a disordered structure related to the exchange between Co and Fe (Mn, or Cr) atoms from the $L2_1$ structure, of which hidden disorder is the so-called $DO3$ disorder (shown in Fig. 1(b)). In general, detection and evaluation of the $DO3$ disorder required special measurement set-ups such as neutron diffraction[21], synchrotron XRD[22-24], and nuclear magnetic resonance[25,26]. In this study, we developed a quantitative analysis technique of the $DO3$ disorder and the other atomic disorders in full-Heusler alloys using a conventional XRD system with Co-$K\alpha$ (0.179 nm) and Cu-$K\alpha$ (0.154 nm) sources. Applying this technique to full-Heusler CFS alloy thin films[8,20] formed by rapid thermal annealing (RTA), we determined the atomic disorders for the exchanges between Co, Fe, and Si atoms in the films quantitatively.

The $L2_1$ structure of full-Heusler $X_2YZ$ alloys consists of eight stacked body-centered cubic (bcc) lattices, and the corner sites of each bcc lattice are occupied by $X$ atoms and the body-centered sites are occupied by $Y$ and $Z$ atoms regularly, as shown in Fig. 1(a). In the case of CFS, $X$, $Y$, and $Z$ are Co, Fe, and Si atoms, respectively. Full-Heusler alloys also have disordered structures referred to as the $B2$ and $A2$ structures. In the $B2$ structure, the $YZ$ sublattice is disordered. For CFS, the exchange between Fe and Si atoms in the $L2_1$ lattice results in the $B2$ structure. The $A2$ structure has no ordered sublattice, i.e., both the $X$ and $YZ$ sublattices are disordered. For CFS, the exchanges among Co, Fe, and Si atoms cause the $A2$ structure. In many cases, these disordered structures are identified by XRD analysis. XRD for full-Heusler alloys can be divided into odd superlattice diffraction, even superlattice diffraction, and fundamental diffraction. The odd superlattice diffraction is defined by the index relation of $h$, $k$, and $l$ = odd numbers, e.g., (111), which can be observed when the $L2_1$ structure is formed. The even superlattice diffraction is defined by $h+k+l=4n+2$, e.g., (200), which appears for the $B2$ and $L2_1$ structures. The fundamental diffraction appears in the case of $h+k+l=4n$, e.g., (220), which is independent of the ordering structures, i.e., this diffraction can be observed even for the $A2$ structure. In general, degrees of order of the $L2_1$ and $B2$ structures can be semi-quantitatively evaluated by the odd and even superlattice diffraction intensities using a hypothesis introduced by Webster[14-20].

However, it is well recognized for full-Heusler CFS (and related $Co_2FeSi_{1-x}Al_x$) alloys that commonly used XRD analysis with an ordinary Cu-$K\alpha$ source cannot detect a disordered structure caused by the exchange between Co and Fe atoms (that is the $DO3$ disorder). This is because the atomic scattering factor ($f_{Co}$) of Co is almost the same as that ($f_{Fe}$) of Fe for the Cu-$K\alpha$ source[18,25], as shown in Fig. 2(a), and thus superlattice diffraction intensities for the $DO3$-disordered structure cannot be distinguished from those for the fully ordered $L2_1$ structure. Since the $DO3$ disorder is a kind of the $A2$ disorder that would strongly degrade the half-metallicity of full-Heusler alloys[11], detection and evaluation of the $DO3$ disorder is highly important. Nevertheless, XRD study for the $DO3$ disorder in full-Heusler alloys has not progressed owing to the above-described difficulty in its detection. It should be noted that although the Webster method is widely used for XRD analysis for full-Heusler CFS alloys[14-20], the amount of the exchange between Co and Fe atoms ($DO3$ disorder) is treated to coincide with that of the exchange between Co and Si atoms[14-17,20]. The similar issue of the $DO3$ disorder has also arisen for other full-Heusler alloys such as CMS and CCA.

The atomic scattering factors ($f_{Co}$ and $f_{Fe}$) of Co and Fe for Co-$K\alpha$ are different due to x-ray anomalous scattering[19], as shown in Fig. 2(b). Therefore, superlattice diffraction intensities for Co-$K\alpha$ can well reflect the $DO3$ disorder. Our quantitative approach using Co-$K\alpha$ and Cu-$K\alpha$ XRD is based on the physical model proposed by Niculescu et al.[27,28] (hereafter, this model is referred to as the NBRB model.) In this model, the disorders in full-Heulser alloys are expressed by three disorder parameters, $\alpha$, $\beta$, and $\gamma$. Table 1 shows the expression of disordered structures using $\alpha$, $\beta$, and $\gamma$, in which $\alpha$ is defined as the number of Fe(Si) atoms occupying on $Z(Y)$ sites per a $Co_2Fe_1Si_1$ unit, which represents the exchange between Fe and Si atoms (i.e., *B*2 disorder), $\beta$ is defined as the number of Co(Si) atoms on $Z(X)$ sites per a $Co_2Fe_1Si_1$ unit, which represents the exchange between Co and Si atoms (hereafter, this exchange is referred to as the *A*2' disorder in order to distinguish it from the *A*2 disorder in the Webster model[14,20,21]), and $\gamma$ is defined as the number of Co(Fe) atoms on $Y(X)$ sites per a $Co_2Fe_1Si_1$ unit, which represents the exchange between Co and Fe atoms (*DO*3 disorder). Note that in the Webster framework, the *A*2' and *DO*3 disorders are treated as the identical disorder (*A*2 disorder) expressed by a single parameter. In the NBRB model, the perfect $L2_1$ structure is expressed by a parameter set of $(\alpha, \beta, \gamma) = (0, 0, 0)$. For the *B*2, *A*2, and *DO*3 disordered structures, maximum $(\alpha, \beta, \gamma)$ values are (1/2, 0, 0), (1/4, 1/2, 1/2), and (0, 0, 2/3), respectively. Using the expression of the disordered structures shown in Table 1, the crystal structure factors[14] $F_{111}$, $F_{200}$, and $F_{220}$ of (111), (200), and (220) diffraction can be written by,

$$F_{111} \propto (1 - 2\alpha - \beta)(f_{Fe} - f_{Si}) + (\gamma - \beta)(f_{Co} - f_{Fe}),  \quad (1)$$

$$F_{200} \propto (1 - 2\beta)(f_{Co} - f_{Si}) + (1 - 2\gamma)(f_{Co} - f_{Fe}),  \quad (2)$$

$$F_{220} \propto 2f_{Co} + f_{Fe} + f_{Si},  \quad (3)$$

where $f_{Co}$, $f_{Fe}$, and $f_{Si}$ are the atomic scattering factors of Co, Fe, and Si, respectively. These values can be taken from Ref. 29. It should be noted that the last terms in Eqs. (1) and (2) include $\gamma$ and have a factor of the difference of $f_{Co}$-$f_{Fe}$.  In the case of Cu-$K\alpha$, there are few differences between $f_{Co}$ and $f_{Fe}$ (shown in Fig. 2(a)).  Therefore, the superlattice diffraction intensities ($I_{111}(\propto F_{111}^2)$ and $I_{200}$ ($\propto F_{200}^2$)) for Cu-$K\alpha$ take constant values even for changes in $\gamma$, since the $\gamma$-related terems in Eqs. (1) and (2) are diminished.  On the other hand, in the case of Co-$K\alpha$, there exist a large difference between $f_{Co}$ and $f_{Fe}$ (shown in Fig. 2(b)), and thus the superlattice diffraction intensities for Co-$K\alpha$ well reflect $\gamma$.  Using both Cu-$K\alpha$ and Co-$K\alpha$ XRD, we can obtain the following four superlattice diffraction intensities: $I_{111}$ and $I_{200}$ for Co-$K\alpha$, and $I_{111}$ and $I_{200}$ for Cu-$K\alpha$.  In order to minimize measurement errors, the three unknown parameters of $\alpha$, $\beta$, and $\gamma$ were determined from these four intensities by the least square method.  In our study, Rigaku SmartLab XRD system with Co-$K\alpha$ and Cu-$K\alpha$ x-ray sources was employed for these superlattice diffraction measurements.

Full-Heusler CFS alloy thin films were prepared by our developed RTA technique[8,20,30].  The CFS films were formed by the RTA-induced silicidation of Co/Fe layers deposited on a Si-on-insulator substrate.  The experimental procedure and fundamental film structure investigated previously were described in Refs. 8 and 20 in detail.  Figures 3(a) and (b) show the in-plane XRD patterns of the CFS films formed at RTA temperature $T_A$ ranging from 650 to 800 °C, measured with the Cu-$K\alpha$ and Co-$K\alpha$ sources, respectively.  Since the CFS thin films were (110)-oriented columnar grain (texture) structures[8,20], the three important diffraction lines of ($1\bar{1}1$), (002), and ($2\bar{2}0$) that are perpendicular to the sample plane can be detected simultaneously by a single scan of the in-plane XRD measurement. The superlattice diffraction lines of CFS($1\bar{1}1$) and CFS(002) were clearly observed for both Co-$K\alpha$ and Cu-$K\alpha$ XRD, implying a highly ordered structure. Using these four diffraction intensities, the disorder parameters were determined.

Red, blue, and green curves in Fig. 4 show the disorder parameters $\alpha$, $\beta$, and $\gamma$ as a function of $T_A$, respectively. All the disorder parameters for the CFS films were relatively small values as the overall feature. Above $T_A$ = 700 °C, they decreased with increasing $T_A$. The film formed at $T_A$ = 800 °C (that is the highest $T_A$ examined here) showed very small values of ($\alpha$, $\beta$, $\gamma$) = (0.06, 0.01, 0.04). The $\alpha$ value was comparable to those of a bulk CFS[31]. These results revealed that all the atoms in the CFS films tend to occupy their original sites of the $L2_1$ structure at higher temperatures. Site occupancies of Co, Fe, and Si atoms on their original sites in the $L2_1$ structure can be easily deduced from the disorder parameters, which were 98 %, 90 %, and 93 %, respectively, for the CFS film formed at $T_A$ = 800 °C. This result revealed that the RTA-formed CFS film has the extremely ordered $L2_1$ structure.

In summary, we developed a new analysis technique for atomic disorder structures in full-Heusler alloys using XRD with Co-$K\alpha$ and Cu-$K\alpha$ sources. The developed technique can quantitatively evaluate all the atomic disorders for the exchanges between *X*, *Y*, and *Z* atoms in full-Heusler $X_2YZ$ alloys with more precise physical model than commonly used Webster's model based on XRD with a single Cu-$K\alpha$ source. In particular, the technique can treat the $DO3$ disorder that cannot be analyzed by Cu-$K\alpha$ XRD. Full-Heusler CFS alloy thin films formed by RTA were analyzed by this method. RTA-temperature dependence of the atomic disorders was investigated.  Site occupancies of Co, Fe, and Si atoms on their original sites were 98 %, 90 %, and 93 %, respectively, for the CFS film formed at RTA temperature of 800 °C.  This result indicated that the RTA-formed CFS film has the extremely ordered $L2_1$ structure.

The authors would like to thank Prof. H. Munekata of Tokyo Institute of Technology, Prof. M. Tanaka and S. Takagi of The University of Tokyo.   In-plane XRD measurements were performed by Rigaku Corporation.

Figure captions

Table 1: Definition of the number of constituent atoms in full-Heulser $Co_2FeSi$ alloy (in unit of a $Co_2Fe_1Si_1$ molecule).

Fig. 1 (a) Schematic illustration of the $L2_1$ structure of full-Heusler alloys. (b) Schematic illustration of the $DO3$ disorder.

Fig. 2 Atomic scattering factors of Co and Fe atoms, $f_{Co}$ and $f_{Fe}$, as a function of $\sin\theta/\lambda$ for (a) Cu-$K\alpha$ and (b) Co-$K\alpha$, where $\theta$ and $\lambda$ are scattering angle of x-ray and incident x-ray wave length, respectively.

Fig. 3 (a) In-plane XRD patterns measured with Cu-$K\alpha$ and (b) Co-$K\alpha$ for $Co_2FeSi$ films formed at $T_A$ ranging from 650 to 800 °C.

Fig. 4 Disorder parameters, $\alpha$ ($B2$ disorder), $\beta$ ($A2'$ disorder), and $\gamma$ ($DO3$ disorder), as a function of $T_A$. Red, blue, and green curves show $\alpha$, $\beta$, and $\gamma$, respectively.

|  | *X* sites | *Y* sites | *Z* sites |
|---|---|---|---|
| **Co** | $2-\gamma-\beta$ | $\gamma$ | $\beta$ |
| **Fe** | $\gamma$ | $1-\alpha-\gamma$ | $\alpha$ |
| **Si** | $\beta$ | $\alpha$ | $1-\alpha-\beta$ |

Table 1 takamura et al.

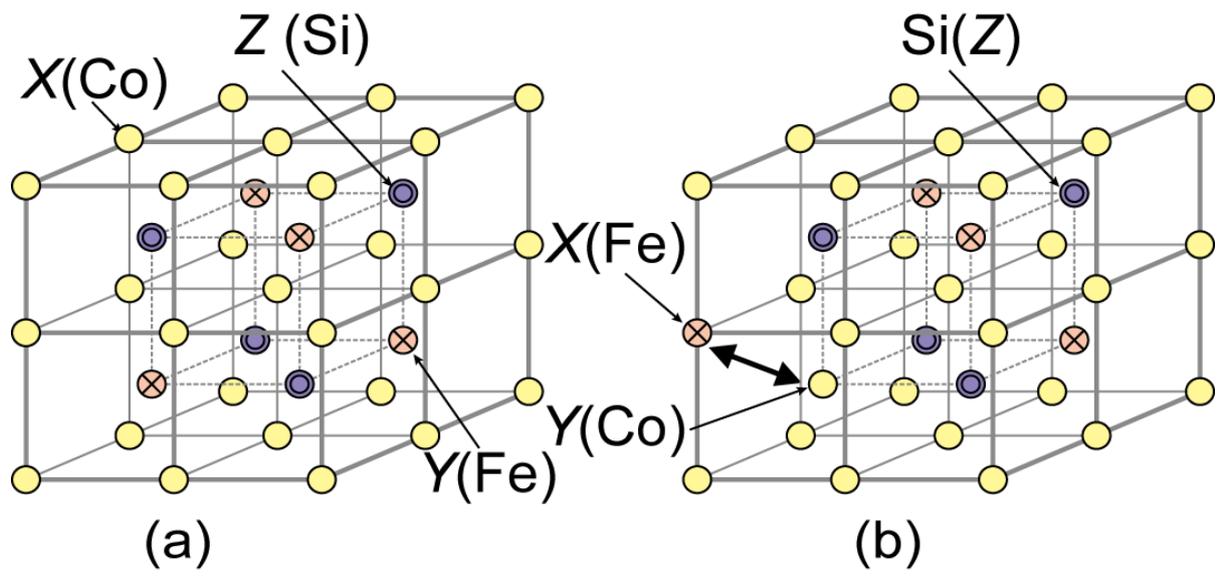

Figure 1 takamura et al.

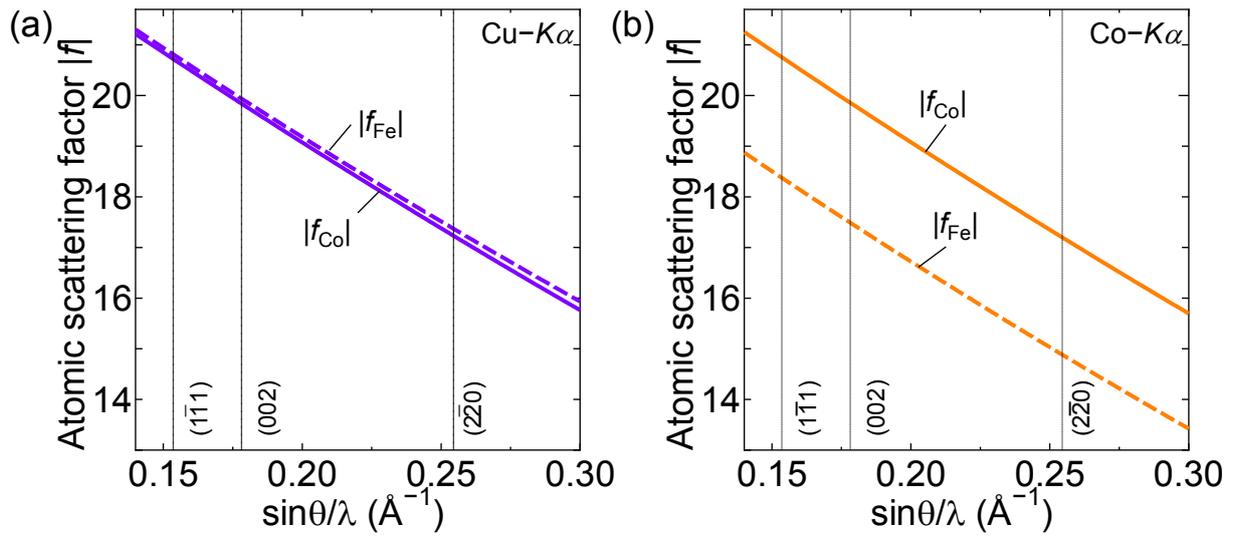

Figure 2 takamura *et al.*

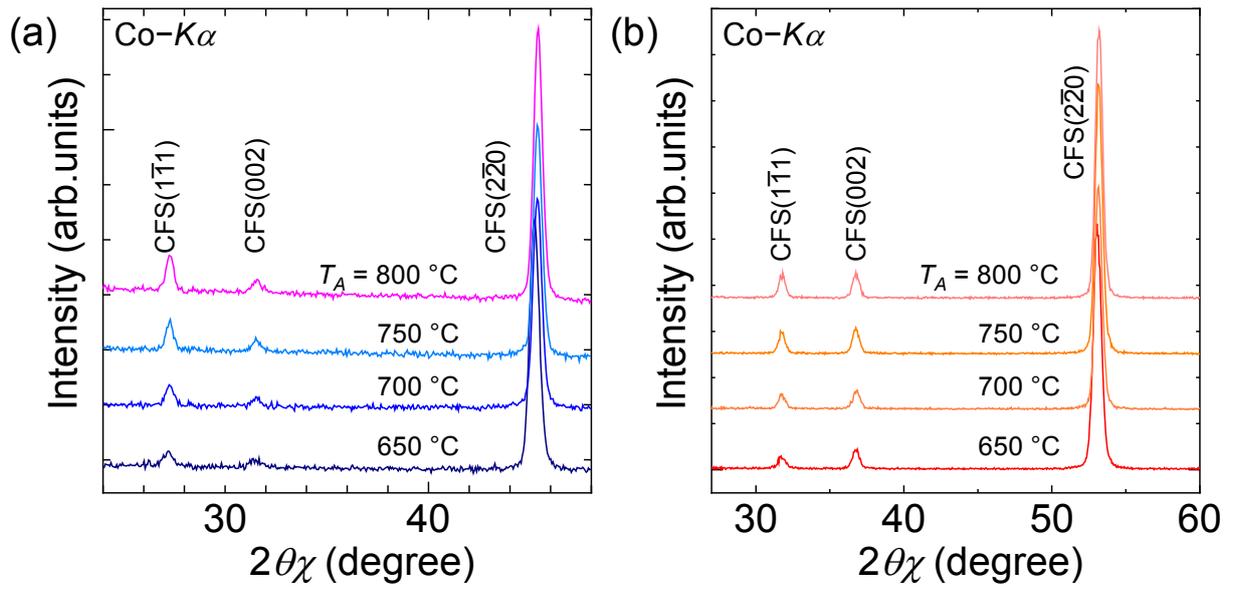

Figure 3 takamura et al.

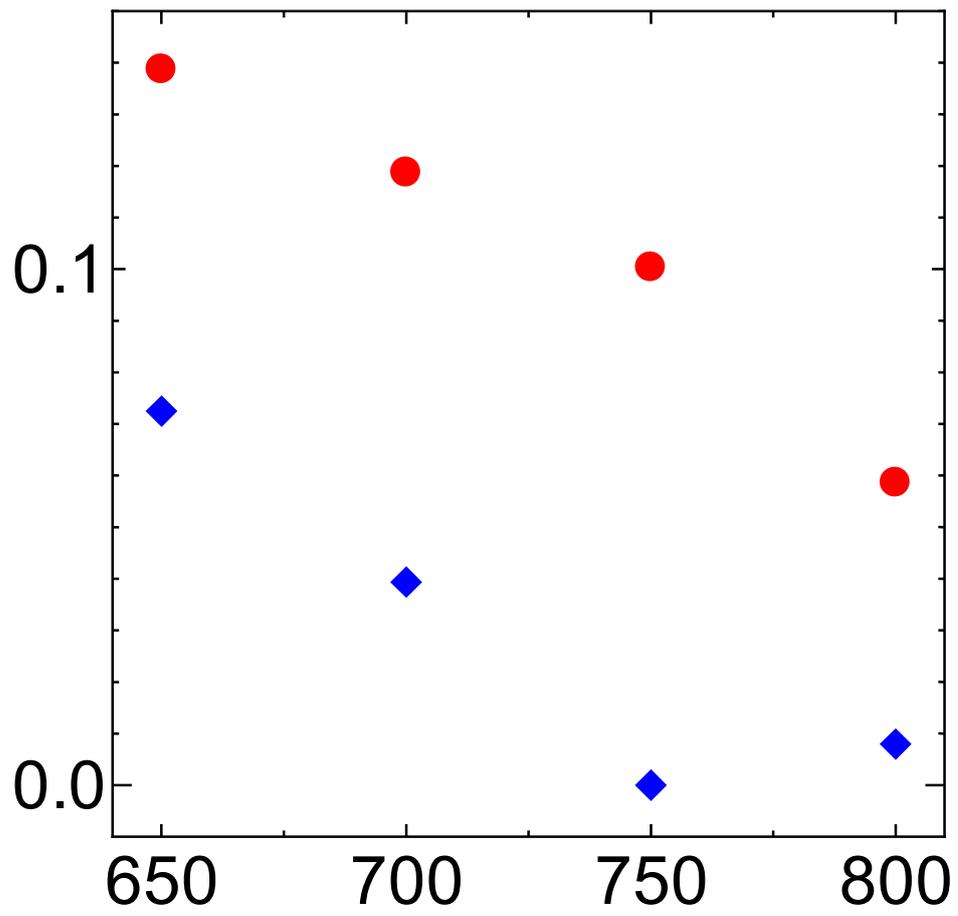

Figure 4 takamura et al.